\documentclass[twocolumn, 10pt]{IEEEtran}

\usepackage{graphicx,amsmath,amsfonts, epsfig, hyperref,
color, nomencl, nicefrac}
\usepackage[thmmarks]{ntheorem}
\usepackage{cite}
\usepackage{psfrag}
\usepackage{multirow}
\usepackage{dsfont}

\newtheorem{example}{Example}[section]
\newtheorem{theorem}{Theorem}
\theoremheaderfont{\normalfont}\theorembodyfont{\normalfont} \theoremstyle{nonumberplain}
\theoremseparator{:}
\theoremsymbol{}

\begin{document}

\title{On the ergodic sum-rate performance of CDD in multi-user systems}

\author{
\authorblockN{Aydin Sezgin\thanks{The work is supported in part by the
  Deutsche Forschungsgemeinschaft (DFG) and by NSF Contract NSF
  DMS-0354674 ONR Contract ONR N00014-02-1-0088-P00006.}, Mohamad Charafeddine  and  Arogyaswami Paulraj} \\
\authorblockA{ Stanford University,\\
 Information Systems Laboratory, \\
 350 Serra Mall, CA 94305-9510, USA, \\
  Tel: +1650 725-6099, Fax: +1650 723-8473 \\
\{sezgin,mohamad,apaulraj\}@stanford.edu}
 }
 \maketitle
\thispagestyle{empty}

\begin{abstract}
The main focus of space-time coding design and analysis for MIMO systems has been so far focused on single-user systems. For
single-user systems, transmit diversity schemes suffer a loss in spectral efficiency if the
receiver is equipped with more than one antenna, making them unsuitable for high rate transmission. One such transmit
diversity scheme is the cyclic delay diversity code (CDD). The advantage of CDD over other diversity schemes such as orthogonal
space-time block codes (OSTBC) is that a code rate of one and delay optimality are achieved independent of the number
of transmit antennas. In this work we analyze the ergodic rate of a multi-user multiple access channel (MAC) with each
user applying such a cyclic delay diversity (CDD) code. We derive closed form expressions for the ergodic sum-rate of
multi-user CDD and compare it with the sum-capacity. We study the ergodic rate region and show that in contrast to what is conventionally known regarding the
single-user case, transmit diversity schemes are viable candidates for high rate transmission in multi-user systems.
Finally, our theoretical findings are illustrated by numerical simulation results.
\end{abstract}

\section{Introduction}
Multiple-input multiple-output or MIMO wireless systems have received a significant amount of interest due to their capability of
dramatically increasing the capacity of a communication link~\cite{Telatar99,FoschiniGans98}. Also, there has been
considerable work on a variety of schemes which exploit multiple antennas at both the transmitter and receiver in order
to obtain spatial diversity, i.e. to improve the reliability of the system such as orthogonal space-time block
codes (OSTBC)~\cite{Alamouti,TarokhJafarkCalder99} and space-time trellis codes~\cite{TarokhSeCa98} (STTC). An simple
example for a STTC is the delay diversity code, first proposed in~\cite{WittnebenSTTC} and also discussed
in~\cite{SeshadriWint94,TarokhSeCa98}, where the data stream on the first transmit antenna is transmitted with delays
on the other antennas. The delay diversity code achieves full diversity but has a disadvantage of a slight rate loss
due to some leading and tailing zeros. In order to avoid this drawback, cyclic delay diversity (CDD) has been proposed
in~\cite{Gore01delay}.

The main advantage of CDD over OSTBC is that a code rate of one is achieved independent of the number of transmit
antennas, whereas OSTBC suffer a rate loss by increasing the number of transmit
antennas~\cite{TarokhJafarkCa99,X.B.Liang2003,WangXiaOSTBC}. The performance of CDD in terms of error probability was
analyzed in\cite{Gore01delay} for frequency flat and frequency selective channels. The diversity-multiplexing tradeoff
for delay diversity was characterized in~\cite{VazeRajanDD}. Based on the assumption that statistical information about
the channel is available at the transmitter, transmit filters employed at the transmitter are optimized
in~\cite{SchoberHehn} in order to minimize the Chernoff bound on the error probability of CDD. The average rate for CDD
by assuming Gaussian as well as PSK/QAM input signals for a point-to-point transmission with $n_T=2$ or $n_T=4$
transmit antennas and $n_R\leq 2$ receive antennas was investigated in terms of Monte-Carlo simulations in~\cite{BauchCDD}.

As noticed in~\cite{BoelcskeiMacStc}, most of the work on space-time coding so far is focused on single-user systems.
For such single-user systems, it was shown that the loss in terms of spectral efficiency for transmit diversity schemes
increases significantly for more than one receive antenna~\cite{Sandhu}, making them inappropriate for high rate
transmission. In this work, by assuming a Gaussian codebook, we analyze the ergodic rate performance of a multiple
access channel (MAC) system with multiple users transmitting their data to the base station by using a CDD. At the
receiver, we apply a joint maximum-likelihood (ML) detector. We compare the achieved ergodic sum-rate with the
sum-capacity of a MAC system assuming that the users do not have channel state information (CSI) and the base station
has full CSI. We show that, if the number of receive antennas $n_R$ is less than the number of users $K$, i.e. $n_R\leq
K$ in the system (which is a reasonable assumption), the loss of the proposed scheme in terms of spectral efficiency is
negligible. This is in strong contrast to single-user systems~\cite{Sandhu,BauchCDD}. Thus, CDD is indeed an attractive
candidate for transmit strategies in MIMO multi-user systems.

The remainder of the paper is organized as follows. In Section~\ref{sec:System}, we introduce the system model and the
construction of the CDD. The performance of CDD in terms of spectral efficiency is then analyzed and compared to the
sum-capacity in section~\ref{sec:PerfAn}, followed by simulations and concluding remarks in Section~\ref{sec:Simu}
and~\ref{sec:Conc}.

Notational conventions are as follows. We will use bold-faced upper case letters to denote matrices, e.g.,
$\mathbf{X}$, with elements $x_{i,j}$; bold-faced lower case letters for column vector, e.g., $\mathbf{x}$, and
light-faced letters for scalar quantities. The superscripts $(\cdot)^T$ and $(\cdot)^H$ denote the transpose and
Hermitian operations, respectively. Finally, the identity matrices and all zero vectors of the required dimensions will
be denoted by $\mathbf{I}$ and $\mathbf{0}^T$. $\mathbb{E}[\cdot]$ will denote the expectation operator.

\section{System model}\label{sec:System}
\begin{figure}
   \begin{center}
      \psfrag{BS}{$BS$}
   \psfrag{MS1}{${MS}_1$}
   \psfrag{MS2}{$MS_2$}
   \psfrag{MS3}{$MS_K$}
\includegraphics[scale=0.6]{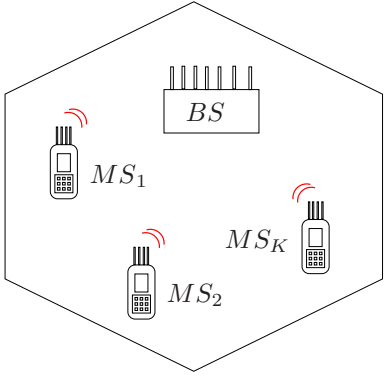}
  \caption{Multiuser MIMO multiple access channel (MAC) with users $MS_1$,\dots, $MS_K$ transmitting data to
  base station (BS).}~\label{fig:MAC}
  \end{center}
\end{figure}

In this work, we consider a multiuser multiple access channel (MAC) with $K$ users each with $n_T$ transmit antennas
and $n_R$ receive antennas at the base station as depicted in Fig.~\ref{fig:MAC}.

 The system model is given as
\begin{align}\label{eq:SysMod}
\mathbf{Y}=\sum_{k=1}^{K}\sqrt{\frac{\mathsf{SNR}}{n_T}}\mathbf{G}_k\mathbf{H}_k^T+\mathbf{N},
\end{align}
where $\mathsf{SNR}$ denotes the signal-to-noise ratio, $\mathbf{G}_k$ is the transmit matrix of size $[T \times n_T]$,
$T$ is the block length and $\mathbf{N}$ ($[T \times n_R]$) is the additive white Gaussian noise (AWGN) matrix
$\mathbf{N} \sim \mathcal{CN}(0,\mathbf{I})$. In this work, we focus on delay optimal codes, in the sense that the decoding is performed after $T$ channel uses with
$n_T=T$\cite{TarokhJafarkCalder99,TarokhJafarkCa99}. The channel between user $k$, $1\leq k \leq K$, and the base
station is modeled by a random zero-mean component channel $\mathbf{H}_k$ ($[n_R \times n_T]$) with identically
independent distributed (iid) complex Gaussian entries, i.e. $\mathbf{H}_k \sim \mathcal{CN}(0,\mathbf{I})$, $1\leq k
\leq K$.
 We assume that each user
$k$ is applying a cyclic delay diversity scheme (for simplicity we skip the user index) , given as
\begin{align}
\mathbf{G}_k(\mathbf{x}_k)=\mathbf{G}(\mathbf{x})=\left[%
\begin{array}{cccc}
  x_1 & x_2 & \cdots & x_T \\
  x_T & x_1 & \cdots & x_{T-1} \\
  x_{T-1} & x_T & \cdots & x_{T-2} \\
  \vdots & \ddots & \ddots & \ddots \\
  x_3 & x_4 & \ddots & x_2 \\
  x_2 & x_3 & \cdots & x_1
\end{array}%
\right]
\end{align}
where $\mathbf{x}_k=[x_1,x_2,\dots,x_T]^T$ is the vector of transmitted symbols of user $k$.
\section{Performance analysis}~\label{sec:PerfAn}
In the following, we analyze the ergodic sum-rate performance achieved with the proposed scheme and compare it to the
ergodic MAC sum-capacity. We start with the case, where only one antenna is available at the base station, i.e.
$n_R=1$, and afterwards generalize it to higher number of receive antennas. The separation of these two cases is
motivated by the fact that the spectral efficiency for transmit diversity schemes in single-user systems with one
receiving antenna is close to (e.g. QSTBC) or even equal to capacity (Alamouti scheme)~\cite{Sandhu}. With more than
one receive antenna, this behavior changes significantly to the disadvantage of the transmit diversity schemes. As we
will see later on, for multi-user systems the picture looks different.

\subsection{$n_R=1$ antenna at the base station}

For $n_R=1$ at the base station, the system equation in~\eqref{eq:SysMod} reduces to
\begin{align}\nonumber
\mathbf{y}=\sum_{k=1}^{K}\sqrt{\frac{\mathsf{SNR}}{n_T}}\mathbf{G}_k\mathbf{h}_k^T+\mathbf{n}
\end{align}
with $\mathbf{h}_k$ of size $[1\times n_T]$, which can be rewritten as
\begin{align}\nonumber
\mathbf{\tilde{y}}=\sum_{k=1}^K\sqrt{\frac{\mathsf{SNR}}{n_T}}\mathbf{\tilde{H}}_k\mathbf{x}_k+\tilde{\mathbf{n}}
=\sqrt{\frac{\mathsf{SNR}}{n_T}}\mathbf{\tilde{H}}\mathbf{x}+\tilde{\mathbf{n}}
\end{align}
with where the circulant matrices $\mathbf{\tilde{H}}_{k}$, $1\leq k \leq K$,
\begin{align}\nonumber
\mathbf{\tilde{H}}_{k}=\left[%
\begin{array}{cccc}
  h_{1}^k & h_{2}^k & \cdots & h_{T}^k \\
  h_{2}^k & h_{3}^k & \cdots & h_{1}^k \\
  h_{3}^k & h_{4}^k & \cdots & h_{2}^k \\
  \vdots & \ddots & \ddots & \ddots \\
  h_{T-1}^k & h_{T}^k & \ddots & h_{T-2}^k \\
  h_{T}^k & h_{1}^k & \cdots & h_{T-1}^k
\end{array}%
\right]
\end{align}
are the effective channels between user $k$ and the base station,
$\mathbf{\tilde{H}}=\left[\mathbf{\tilde{H}}_{1},\mathbf{\tilde{H}}_{2},\dots,\mathbf{\tilde{H}}_{K}\right]$ and
$\mathbf{x}=[\mathbf{x}_1^T, \mathbf{x}_2^T, \dots, \mathbf{x}_K^T]^T$ is obtained by stacking the transmit signals of
the users into one large vector. The instantaneous  rate achievable with the circulant code is then given as
\begin{align}
I_c &=\frac{1}{T}\log_2\det\left(\mathbf{I}_T+\frac{\mathsf{SNR}}{n_T}\mathbf{\tilde{H}}\mathbf{\tilde{H}}^H\right) \nonumber \\
&
=\frac{1}{T}\log_2\det\left(\mathbf{I}_T+\frac{\mathsf{SNR}}{n_T}\sum_{k=1}^K\mathbf{\tilde{H}}_k\mathbf{\tilde{H}}_k^H\right).\nonumber
\end{align}
Since the $\mathbf{\tilde{H}}_k$, $1\leq k \leq K$, are all circulant matrices, they can be simultaneously diagonalized
by the unitary DFT matrix $\mathbf{D}$~\cite{Davis}, i.e. $\mathbf{\tilde{H}}_k\mathbf{\tilde{H}}_k^H=\mathbf{D} n_T \mathbf{\Lambda}_k\mathbf{D}^H$. Therefore, $I_c$ results in
\begin{align}
I_c & =\frac{1}{T}\log_2\det\left(\mathbf{I}_T+\frac{\mathsf{SNR}}{n_T}\sum_{k=1}^K\mathbf{D} n_T \mathbf{\Lambda}_k\mathbf{D}^H\right),\nonumber\\
& =\frac{1}{T}\log_2\det\left(\mathbf{I}_T+\frac{\mathsf{SNR}}{n_T}n_T\mathrm{diag}
\left(\lambda_1,\lambda_2,\dots,\lambda_T\right)\right),\nonumber
\end{align}
where the $\lambda_t$ are i.i.d. chi-square distributed random variables with $2K$ degrees of freedom, i.e. $\lambda_t
\sim \chi_{2K}^2$. For the ergodic sum-rate performance given as $R_c=\mathbb{E}[I_c]$, we thus have
\begin{align}\nonumber
R_c=\frac{1}{T}\mathbb{E}\left[\sum_{t=1}^T\log_2\left(1+\mathsf{SNR}\lambda_t\right)\right]
=\mathbb{E}\left[\log_2\left(1+\mathsf{SNR}\lambda\right)\right],
\end{align}
where $\lambda \sim \chi_{2K}^2$. In order to derive a lower bound on the above expression, we make use of the fact
that $\log_2\left(1+a e^x\right)$ is a convex function with $a>0$, thus applying Jensen's inequality results in
\begin{align}
R_c \geq \log_2\left(1+\mathsf{SNR}\exp\left(\mathbb{E}\left[\log_2\left(\lambda\right)\right]\right)\right).\nonumber
\end{align}
From~\cite{OymanNBPaulraj,GrantUppBound} we know that
\begin{align}\label{eq:EWChi2DiGam}
\mathbb{E}\left[\log_2\left(\lambda\right)\right]=\psi(K)=\sum_{k=1}^{K-1}\frac{1}{k} -\gamma,
\end{align}
where $\psi(\cdot)$ is the digamma or psi function~\cite[p.943, eq.8.360]{Gradshteyn} and $\gamma \approx 0.577$ is
Euler's constant. Thus, for $n_R=1$ receive antenna at the base station, the average sum-rate performance of the
proposed scheme is lower bounded by
\begin{align}\label{eq:LbCddRateNr1}
R_c \geq \log_2\left(1+\mathsf{SNR}\exp\left(\sum_{k=1}^{K-1}\frac{1}{k} -\gamma\right)\right).
\end{align}
In contrast, note that the sum-capacity of the multiuser MAC system can be lower bounded by
\begin{align}\label{eq:LbSumCapNr1}
C \geq \log_2\left(1+\frac{\mathsf{SNR}}{n_T}\exp\left(\sum_{k=1}^{n_TK-1}\frac{1}{k} -\gamma\right)\right).
\end{align}
The result in~\eqref{eq:LbSumCapNr1} is obtained by applying the techniques in~\cite{OymanNBPaulraj} and the fact that
for no CSIT the ergodic sum capacity of a K users MAC channel, where each user has $n_T$ transmit antennas, is
equivalent to the ergodic capacity of a single-user system with $Kn_T$ transmit antennas~\cite[Proposition
1]{RheeCioffi},\cite{Telatar99}. By comparing~\eqref{eq:LbCddRateNr1} and~\eqref{eq:LbSumCapNr1}, we observe that the
difference between the ergodic sum-rate achieved with CDD and the sum-capacity, i.e. $C-R_c$ converges to
\begin{align}
& \frac{1}{\ln(2)}\sum_{k=K}^{n_TK-1}\frac{1}{k}-\ln(n_T) \nonumber\\
& =\frac{1}{K\ln(2)}\left(1+K\left(\sum_{k=K+1}^{n_TK-1}\frac{1}{k}-\ln(n_T)\right)\right)\label{eq:RateDiff}
\end{align}
for high $\mathsf{SNR}$. This result is restricted to high $\mathsf{SNR}$, where the lower bounds in~\eqref{eq:LbCddRateNr1} and~\eqref{eq:LbSumCapNr1} are tight and thus can be used for computing the difference $C-R_c$.  Note that $\sum_{k=K+1}^{n_TK-1}\frac{1}{k}$ is a nondecreasing function in $K$ and $n_T$.
Further, note that
\begin{align}
& \lim_{K\rightarrow \infty} \sum_{k=K+1}^{n_TK-1}\frac{1}{k}=\lim_{K\rightarrow \infty}
\sum_{k=1}^{n_TK-1}\frac{1}{k}-\sum_{k=1}^{K}\frac{1}{k} \nonumber\\
&=\lim_{K\rightarrow \infty} \psi(n_TK)-\psi(K+1)=\ln(n_T). \label{eq:LimPsiFunc}
\end{align}
Thus, for $n_R=1$ the difference in~\eqref{eq:RateDiff}, i.e. $C-R_c$, is upper bounded by
\begin{align}
C-R_c\leq \frac{1}{K \ln(2)}
\end{align}
for high SNR, i.e. the more the number of users the less is the difference between the actual capacity and the rate
achieved with CDD.

\subsection{Arbitrary number of receive antennas $n_R$}

For higher number of receive antennas the effective (sometimes also referred to as equivalent) channel to each user $k$
is given as
\begin{align}
\mathbf{\tilde{H}}_k=\left[\mathbf{\tilde{H}}_{k,1}^T, \mathbf{\tilde{H}}_{k,2}^T, \dots,
\mathbf{\tilde{H}}_{k,n_R}^T\right]^T\nonumber
\end{align}
where
\begin{align}\nonumber
\mathbf{\tilde{H}}_{k,i}=\left[%
\begin{array}{cccc}
  h_{i,1}^k & h_{i,2}^k & \cdots & h_{i,T}^k \\
  h_{i,2}^k & h_{i,3}^k & \cdots & h_{i,1}^k \\
  h_{i,3}^k & h_{i,4}^k & \cdots & h_{i,2}^k \\
  \vdots & \ddots & \ddots & \ddots \\
  h_{i,T-1}^k & h_{i,T}^k & \ddots & h_{i,T-2}^k \\
  h_{i,T}^k & h_{i,1}^k & \cdots & h_{i,T-1}^k
\end{array}%
\right]
\end{align}
with
\begin{align}
\mathbf{\tilde{H}}=\left[\mathbf{\tilde{H}}_{1},\mathbf{\tilde{H}}_{2},\dots,\mathbf{\tilde{H}}_{K}\right]
\end{align}
being a $[n_Rn_T \times n_TK]$ matrix. Thus, $\mathbf{\tilde{H}}$ is a block matrix with circulant blocks. Similarly to
the single receive antenna case, the instantaneous rate with arbitrary $n_T$, $n_R$ and $K$ users achievable with the
circulant code is then given as
\begin{align}\label{eq:RateGener}
I_c &=\frac{1}{T}\log_2\det\left(\mathbf{I}_{n_Rn_T}+\frac{\mathsf{SNR}}{n_T}\mathbf{\tilde{H}}\mathbf{\tilde{H}}^H\right).
\end{align}
We are now able to state the following theorem.
\begin{theorem}~\label{theo:SumRate}
The ergodic-sum rate of a multi-user MAC system with K users each equipped with $n_T$ transmit antennas applying a CDD
and $n_R$ receive antennas at the base station is lower bounded by
\begin{align}\label{eq:RateCDDNRGen}
R_c \geq \sum_{l=1}^{L}\log_2\left(1+\mathsf{SNR} \exp\left(\sum_{k=1}^{M-l}\frac{1}{k}-\gamma\right)\right),
\end{align}
with $L=\min(n_R,K)$ and  $M=\max(n_R,K)$.
\end{theorem}
\proof: The proof is given in the Appendix.

In comparison, using similar techniques the sum-capacity of the multiuser MAC system with $K$ users, each with $n_T$ transmit and $n_R$ receive
antennas can be lower bounded by
\begin{align}\label{eq:LBCapGen}
C \geq \sum_{l=1}^{\tilde{L}}\log_2\left(1+\frac{\mathsf{SNR}}{n_T}\exp\left(\sum_{k=1}^{\tilde{M}-l}\frac{1}{k}
-\gamma\right)\right)
\end{align}
with $\tilde{L}=\min(n_R,n_TK)$,  $\tilde{M}=\max(n_R,n_TK)$.

By applying Jensens inequality, the lower bound on $R_c$ in~\eqref{eq:RateCDDNRGen} and the lower bound for $C$ in~\eqref{eq:LBCapGen}
may be bounded from below by
\begin{align}\label{eq:RcBound2}
R_c \geq L\log_2\left(1+\mathsf{SNR}
\exp\left(\frac{1}{L}\sum_{l=1}^{L}\sum_{k=1}^{M-l}\frac{1}{k}-\gamma\right)\right)
\end{align}
and
\begin{align}\label{eq:CBound2}
C \geq \tilde{L}\log_2\left(1+\frac{\mathsf{SNR}}{n_T}\exp\left(\frac{1}
{\tilde{L}}\sum_{l=1}^{\tilde{L}}\sum_{k=1}^{\tilde{M}-l}\frac{1}{k} -\gamma\right)\right),
\end{align}
respectively. These new lower bounds get tight for high SNR~\cite{OymanNBPaulraj}. Assume that $n_R \leq K$, which is a
reasonable assumption. Then, applying similar steps as in~\eqref{eq:RateDiff}-~\eqref{eq:LimPsiFunc}, using~\eqref{eq:RcBound2} and~\eqref{eq:CBound2} the difference
$C-R_c$ may be upper bounded by
\begin{align}\nonumber
C-R_c \leq &
\frac{1}{\ln(2)}\sum_{l=1}^{n_R}\frac{1}{K-l+1}\Bigg(1+(K-l+1) \\
& \times \ln\left(\frac{1}{n_T}+\frac{(n_T-1)K}{n_T(K-l+1)}\right)\Bigg)
\end{align}
for high SNR. Thus, increasing $K$ reduces the difference $C-R_c$, while increasing $n_R$ increases the difference $C-R_c$.

In addition to the lower bound, an upper bound is also derived following the approach of~\cite{GrantUppBound}. Applying
Jensen's inequality to~\eqref{eq:BDMwMIMO} results in
\begin{align}
R_c \leq \log_2\mathbb{E}\left[\det\left(\mathbf{I}+\mathsf{SNR}\mathbf{H}'_1\mathbf{H}'^H_1\right)\right]\nonumber
\end{align}
due to the concavity of the logarithm.
With~\cite[Theorem A.4, eq.(A.11)]{GrantUppBound}, we arrive at
\begin{align}\label{eq:BoundGrant}
    R_c \leq \log_2\left(\sum_{i=0}^L \binom{L}{i}\frac{M!}{(M-i)!}\mathsf{SNR}^i\right).
\end{align}
In the following section, our theoretical results are illustrated by numerical simulations.

\section{Simulations}\label{sec:Simu}

\subsection{Single-User case}
\begin{figure}
\begin{center}
\includegraphics[scale=0.5]{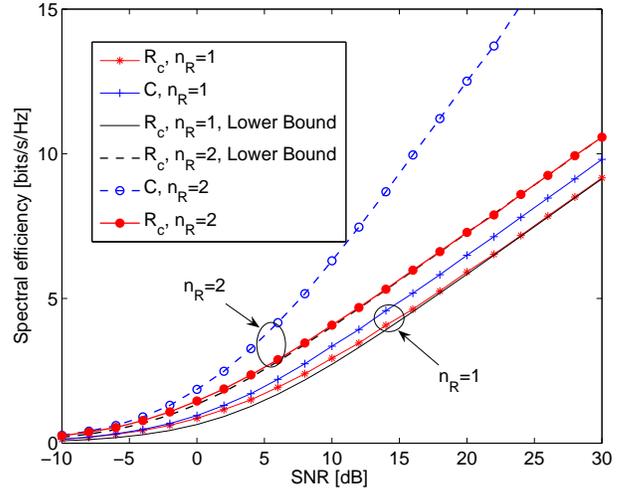}%{UA-5.eps}
\end{center}
\caption{Single-user ergodic rate of the CDD and capacity. $n_R=1,2$ and $n_T=4$. Lower bounds obtained
from~\eqref{eq:RateCDDNRGen} are also depicted.} \label{fig:CapSU}
\end{figure}

In Fig.~\ref{fig:CapSU}, the ergodic rate $R_c$ achievable with the CDD and the ergodic capacity for a single-user
system are depicted with $n_T=4$ transmit and $n_R=1$ and $n_R=2$ receive antennas, respectively. In addition to this,
the lower bound on $R_c$ in~\eqref{eq:RateCDDNRGen} derived in the previous section is also depicted. From the figure,
we observe that for $n_R=1$, $R_c$ is close to the capacity. By increasing $n_R$, we observe that the loss increases
significantly. The lower bound on $R_c$ confirms this behavior and verifies the numerical results obtained
in~\cite{Sandhu,BauchCDD} that transmit diversity schemes perform poorly for $n_R>1$ in single-user systems.

\subsection{Multi-User case}
\begin{figure}
\begin{center}
\includegraphics[scale=0.5]{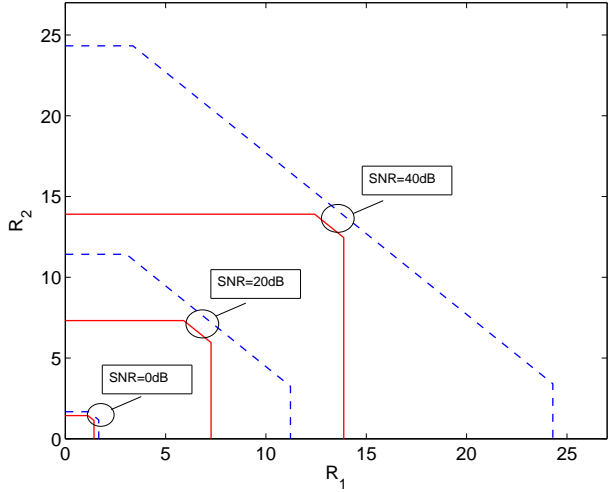}%{UA-5.eps}
\end{center}
\caption{Ergodic Capacity Region (dashes lines) and ergodic rate region of the CDD (solid lines) with $K=2$ users,
$n_R=2$ and $n_T=2$ for $\mathsf{SNR}=0$dB, $\mathsf{SNR}=20$dB and $\mathsf{SNR}=40$dB.} \label{fig:CapRegion}
\end{figure}

In Fig.~\ref{fig:CapRegion}, the ergodic capacity region of a symmetric system with $K=2$ users each equipped with $n_T=2$
transmit antennas and a base station with $n_R=2$ receive antennas is depicted for three different $\mathsf{SNR}$
values ranging from $\mathsf{SNR}=0$~dB to $\mathsf{SNR}=40$~dB. In addition to that, the ergodic rate region
achievable with CDD is depicted. The corner points of the regions can be achieved by a successive cancelation decoder.
Note that the capacity and the maximum rate achievable with CDD of the point-to point link with the other user absent
from the system (or equivalently completely canceled) are given by
\begin{align}
R_c^{1} &\leq\mathbb{E}\log_2\left[\det\left(\mathbf{I}+\mathsf{SNR}\mathbf{h}'_{1,1}\mathbf{h}'^H_{1,1}\right)\right] \nonumber\\
C_1 &\leq \mathbb{E}\log_2\left[\det\left(\mathbf{I}+\frac{\mathsf{SNR}}{n_T}\mathbf{H}_1\mathbf{H}^H_1\right)\right]
\end{align}
and
\begin{align}
R_c^{2} &\leq \mathbb{E}\log_2\left[\det\left(\mathbf{I}+\mathsf{SNR}\mathbf{h}'_{1,2}\mathbf{h}'^H_{1,2}\right)\right] \nonumber\\
C_2 &\leq \mathbb{E}\log_2\left[\det\left(\mathbf{I}+\frac{\mathsf{SNR}}{n_T}\mathbf{H}_2\mathbf{H}^H_2\right)\right],
\end{align}
respectively, where $\mathbf{h}'_{1,i}$ ($[n_R\times 1]$) is the $i$th column of the $n_R \times K$ matrix
$\mathbf{H}'_1$ in~\eqref{eq:BDMwMIMO}. The rate of the individual user cannot exceed these bounds, which are referred
to as single-user bounds with $R_c^1+R_c^2=C$ and $C_1+C_2=C$. From the figure, we observe that the gap between the single-user bounds, e.g. $C_1-R_c^{1}$,
is significantly large. An interesting observation is, however, that the corner points of the capacity region are to
the left or below the corner points of the ergodic rate region of CDD. Thus, at the so called Pareto optimal
part~\cite{TseBook} of the rate region which is the time-sharing part between the two corner points, the gap between the symmetric CDD rate region to the capacity region
is dramatically reduced. The Pareto optimal part contains all the optimal operating points of the
channel~\cite[Ch.6,p.231]{TseBook}.

\begin{figure}
\begin{center}
\includegraphics[scale=0.5]{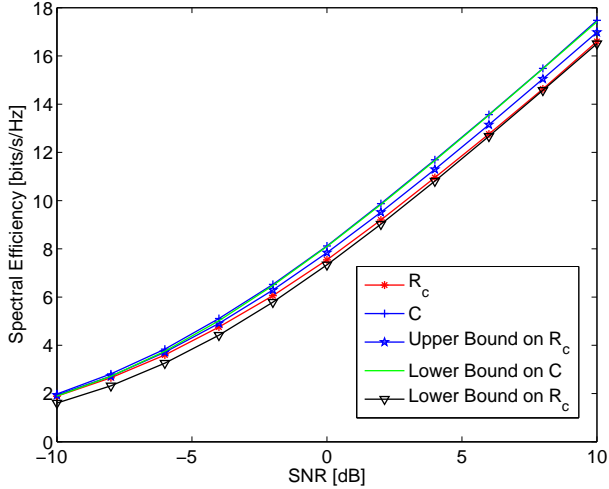}%{UA-5.eps}
\end{center}
\caption{Ergodic sum-rate of the CDD and sum-capacity with $K=6$ users, $n_R=3$ and $n_T=3$. Upper and lower bounds
obtained from~\eqref{eq:BoundGrant} and~\eqref{eq:LBCapGen},\eqref{eq:RateCDDNRGen}, respectively, are also depicted.}
\label{fig:SumCapMU}
\end{figure}

In Fig.~\ref{fig:SumCapMU}, we depict for a system with $K=6$ users and $n_R=n_R=3$ antennas from top to bottom (at the most right): the ergodic capacity, the lower bound on
the ergodic capacity in~\eqref{eq:LBCapGen} (very tight), the upper bound on the ergodic sum-rate of the CDD $R_c$
in~\eqref{eq:BoundGrant}, $R_c$, and the lower bound on $R_c$ in~\eqref{eq:RateCDDNRGen} discussed in the previous
sections. Differently from the single-user system, most of the ergodic capacity is obtained by the CDD despite multiple
receiving antennas at the base station.  Both the lower and upper bound on $R_c$ track the ergodic sum-rate performance
of the CDD quite well.

\section{Conclusion}\label{sec:Conc}
Research in the area of space-time design and analysis for MIMO systems has traditionally focused on
single-users scenarios. For these systems, transmit diversity schemes suffer a loss in spectral efficiency
when the receiver has more than one antenna, making them unsuitable for high rate transmission. Here, we
focused on the performance of cyclic delay diversity (CDD) codes. We analyzed the ergodic sum-rate and studied the
ergodic rate region achievable by using the CDD for each user in a multi-user multiple access channel (MAC). We derived tight
closed form expressions for the ergodic sum-rate of multi-user CDD and compared it with the sum-capacity. We showed
that in contrast to the single-user case, transmit diversity schemes are viable candidates for high rate transmission
in multi-user systems when the number of users exceeds the number of antennas at the base.

\appendix

\subsection{Proof of Theorem~\ref{theo:SumRate}}

\begin{proof}
 The expression in~\eqref{eq:RateGener} may be rewritten as
\begin{align}\label{eq:BMwDB}
I_c=\frac{1}{T}\log_2\det\Big(\mathbf{I}+\frac{\mathsf{SNR}}{n_T}\underbrace{\left( \mathbf{I}_{n_R}\otimes
\mathbf{D}\right)\mathbf{\tilde{H}}\mathbf{\tilde{H}}^H\left( \mathbf{I}_{n_R}\otimes
\mathbf{D}\right)^H}_{n_T\mathbf{\hat{H}}\mathbf{\hat{H}}^H}\Big)
\end{align}
where $\mathbf{D}$ is the unitary DFT-matrix of appropriate size of $[n_T\times n_T]$ with $\mathbf{D}\mathbf{D}^H=\mathbf{I}$ and $\otimes$
denotes the Kronecker product between two matrices. According to~\cite[p.183]{Davis}, the matrix $\mathbf{\hat{H}}\mathbf{\hat{H}}^H$ is
a $[n_Rn_T\times n_Rn_T]$ block matrix with diagonal blocks of size $[n_T\times n_T]$.
 The non-zero entries of the matrix $\mathbf{\hat{H}}$ are each
 $\mathcal{CN}(0,1)$ distributed.

Let us now multiply the matrix $\mathbf{\hat{H}}\mathbf{\hat{H}}^H$ in~\eqref{eq:BMwDB} from right and left with a $[n_Rn_T\times n_Rn_T]$
permutation matrix $\mathbf{P}$ and $\mathbf{P}^H$, respectively. It holds that $\mathbf{P}\mathbf{P}^H=\mathbf{I}$.
\begin{example}
For $n_T=4$, $n_R=2$ and $K=2$, $\mathbf{P}$ is given as
\begin{align}\nonumber
\mathbf{P}=\left[%
\begin{array}{cccccccc}
  1 & 0 & 0 & 0 & 0 & 0 & 0 & 0 \\
  0 & 0 & 0 & 1 & 0 & 0 & 0 & 0 \\
  0 & 0 & 0 & 0 & 1 & 0 & 0 & 0 \\
  0 & 0 & 0 & 0 & 0 & 0 & 1 & 0 \\
  0 & 1 & 0 & 0 & 0 & 0 & 0 & 0 \\
  0 & 0 & 1 & 0 & 0 & 0 & 0 & 0 \\
  0 & 0 & 0 & 0 & 0 & 1 & 0 & 0 \\
  0 & 0 & 0 & 0 & 0 & 0 & 0 & 1 \\
\end{array}%
\right].
\end{align}
\end{example}
Thus, we have
\begin{align}
I_c
&=\frac{1}{T}\log_2\det\left(\mathbf{I}+\mathsf{SNR}\underbrace{\mathbf{P}^H\mathbf{\hat{H}}\mathbf{\hat{H}}^H\mathbf{P}}_{\mathbf{H}'\mathbf{H}'^H}\right)\nonumber\\
&=\frac{1}{T}\sum_{i=1}^{n_T}\log_2\det\left(\mathbf{I}_{n_R}+\mathsf{SNR}\mathbf{H}'_i\mathbf{H}'^H_i\right)\label{eq:BDMwMIMOInst},
\end{align}
where $\mathbf{H}'=\mathrm{diag}\left(\mathbf{H}'_1, \mathbf{H}'_2,\dots, \mathbf{H}'_{n_T}\right)$ is a block diagonal
matrix, with $n_T$ blocks of size $[n_R \times K]$ and each block is $\mathcal{CN}(\mathbf{0},\mathbf{I})$ distributed.
Eq.~\eqref{eq:BDMwMIMOInst} can be interpreted as the sum-rate of $n_T$ parallel MIMO channels each with $n_R$ receive
and $K$ transmit antennas, divided by the block length of $T$. Taking the expectation of~\eqref{eq:BDMwMIMOInst} with
$T=n_T$ results in
\begin{align}\label{eq:BDMwMIMO}
R_c &=\mathbb{E}\log_2\det\left(\mathbf{I}+\mathsf{SNR}\mathbf{H}'_1\mathbf{H}'^H_1\right).
\end{align}
Following the same approach as of~\cite{OymanNBPaulraj}, and using the lower bound on mutual information
from~\cite[eq.(12)]{FoschiniGans98} that
\begin{align}\nonumber
I \geq \sum_{j=1}^{\min(n_R,K)} \log_2 \left(1+\mathsf{SNR} X_j\right),
\end{align}
where the $X_j\sim \chi_{2(\max(n_R,K)-j+1)}^2$, we apply Jensen's inequality with
\begin{align}\nonumber
\mathbb{E}\left[\log_2\left(1+a\exp(x)\right)\right]\geq \log_2\left(1+a\exp(\mathbb{E}\left[x\right])\right)
\end{align} for $a>0$ to~\eqref{eq:BDMwMIMO}. Furthermore, by using~\eqref{eq:EWChi2DiGam} we arrive at the lower bound for $R_c$ given in~\eqref{eq:RateCDDNRGen}.
\end{proof}

 %\tiny{
%\bibliographystyle{IEEE}
%  \bibliography{..//..//..//Bibtexreferenzen//Literatur}
%   %\bibliography{Literatur}
%}

\end{document}